\documentstyle[12pt,graphics,axodraw,setspace,amssymb]{article}
\input epsf

\begin{document}
\baselineskip 24pt

\newcommand{\be}{\begin{equation}}
\newcommand{\ee}{\end{equation}}

\newcommand{\bea}{\begin{eqnarray}}
\newcommand{\eea}{\end{eqnarray}}

\newcommand{\ol}{\overline}
\newcommand{\papertitle}
{LFV and FCNC in a large $\tan\beta$ 
SUSY-seesaw\footnote{Based on a talk given by J.Parry at the 
Mini-workshop on Particle Physics Phenomenology, 
Tainan, R.O.C., June 5th-6th 2006 and the article \cite{Parry:2005fp}}}

\newcommand{\paperauthor}
{\bf J.K.Parry\footnote{e-mail: jkparry@cycu.edu.tw}}

\newcommand{\paperaddress}
{
Department of Physics, 
Chung Yuan Christian University,\\
Chung-Li, Taiwan 320, Republic of China
}

\newcommand{\paperabstract}
{
Realistic predictions are made for lepton flavour violating
and flavour changing neutral current decays in the 
large $\tan\beta$ regime of a SUSY-seesaw model. 
Lepton flavour violating neutral Higgs decays are discussed within
this framework along with the highly constraining charged lepton
decays $\ell_i\to \ell_j \gamma$. In the $b\!-\!s$ 
system the important constraint
from $b\to s \gamma$ is studied in conjunction with the rare FCNC
$B_s\to\mu\mu(\tau\mu)$ decays and the recently measured $\Delta M_s$.
}

\begin{spacing}{0.9}

\begin{center}
{\large{\bf \papertitle}}
\\ \bigskip \paperauthor \\ \mbox{} \\ {\it \paperaddress} \\ 
\bigskip 
\bigskip 
{\bf Abstract} 
\bigskip 
\end{center} 
\begin{center}
\begin{minipage}[t]{15cm}
\paperabstract
\begin{flushleft}
\today
\end{flushleft}
\end{minipage}
\end{center}
\section{Introduction}

Supersymmetric theories contain a number of possible sources of 
lepton flavour violation. As a result, the bounds on rates of LFV and FCNC
are particularly restricting upon the SUSY parameter space.
Even in the case of minimal supergravity(mSUGRA), where soft SUSY breaking
masses and trilinear couplings are flavour diagonal at the GUT scale,
the presence of right-handed neutrinos and the PMNS mixings are 
enough to radiatively induce
large LFV rates. It is these LFV 
rates, induced by renormalisation group(RG) 
running that we wish to study in the present work. 

A large value of $\tan\beta$ enhances LFV and FCNC effects 
in SUSY theories and 
it is interesting to study them together within a SUSY-GUT scenario.
In the large $\tan\beta$ limit it has been noted that Higgs 
mediated contributions to FCNC decays may be dominant.
The most interesting example of Higgs mediated decays is that
of $B^{0}_{s}\to\mu^+\mu^-$\cite{Bsmm_papers,Parry:2005fp}
in the MSSM. This decay is proportional to $\tan^{6}\beta$ and so is 
greatly enhanced by large values of $\tan\beta$. 
If the Higgs boson is very light then this rate 
could become very exciting. 
Another interesting case is the Higgs mediated contribution
to the recently measured $B_s^0\!-\!\bar{B}_s^0$ mixing at the Tevatron.
Along with improved bounds on leptonic $B_s$ decays these
measurements add a complimentary $b-s$ constraint to that of $b\to s\gamma$.

We propose to study the rates of lepton flavour violating decays of
$\tau$, $B_{s}^{0}$ and MSSM neutral 
Higgs bosons in a SUSY-seesaw model with bi-large neutrino mixing. 
From this framework we hope to make general 
predictions for some of the interesting lepton flavour violating decays
mentioned above and also to study their correlations.
The present work differs from previous studies 
in that it makes use of a top-down $\chi^2$ 
analysis of a Grand Unified
SUSY model. Also we are studying lepton flavour violating Higgs decays
in conjunction with Higgs mediated decays, such as $B_s\to\mu^+\mu^-$.

\section{A large $\tan\beta$ SUSY-seesaw model}\label{theory}

In the present work we shall study the MSSM$+\nu_R$ 
constrained at the GUT scale by minimal $SU(5)$ unification. 
In Minimal $SU(5)$ the matter superfields of the MSSM$+\nu_R$ are 
contained within the representations;
${\bf 10}=(Q,\,U^c,\,E^c)$, and 
${\bf \bar{5}}=(L,\,D^c)$ and ${\bf 1}=(N^c)$.
In addition there are ${\bf 5}=\left(H_u,\,H_c \right)$ 
and ${\bf \bar{5}}=\left(H_d,\,\bar{H}_c \right)$ Higgs representations.
Using these matter superfields we can construct the Yukawa section of the 
superpotential as follows,
\be
W_Y^{SU(5)} = \frac{1}{8}\,Y_{u}^{ij} \, {\bf 10}_i {\bf 10}_j {\bf H} 
+ Y_{d}^{ij} \, {\bf 10}_i {\bf \bar{5}}_j {\bf \bar{H}}
+Y_{\nu}^{ij} \, {\bf \bar{5}}_i {\bf 1}_j {\bf H} 
+ \frac{1}{2}\,M_{R}^{ij} \, {\bf 1}_{i}{\bf 1}_{j}. \label{Wy}
\ee
The first term of eq.~(\ref{Wy}) gives rise to a 
symmetric up-quark Yukawa coupling $Y_{u}$ and 
the second term provides both down-quark
and charged lepton Yukawa couplings. As a result we
have the GUT scale relation, $Y_{e}=Y_{d}^{\sf T}$.
The final two terms 
of eq.~(\ref{Wy}) are responsible for the 
neutrino Yukawa coupling, $Y_{\nu}$ and the right-handed neutrino
Majorana mass, $M_{R}$ which combine to produce the 
light neutrino mass matrix via the type I seesaw mechanism,
\be
m_{LL}= -\, v_u^2\,\,Y_{\nu}\,M_R^{-1}\,Y_{\nu}^{\sf T}\label{seesaw}.
\ee

We begin by rotating the ${\bf 10}$ and ${\bf \bar{5}}$ representations 
such that $Y_{d}$ is diagonal.
As a result $Y_{u}$ is solely responsible
for the CKM mixing. The lepton sector is more complicated due to the 
structure of the see-saw mass matrix eq.~(\ref{seesaw}).
Hence, in this basis we have,
\bea
Y_u&=&V_{\sf CKM}^{\sf T}\,Y_u^{\sf diag}\,V_{\sf CKM}\nonumber    \\
Y_d&=&Y_d^{\sf diag}\nonumber    \\
Y_{\nu}&=&U\,Y_{\nu}^{\sf diag}\label{Yuk}    \\
M_R&=&W\;M_R^{\sf diag}\;W^{\sf T}\nonumber .
\eea
Here the singlet neutrino has been rotated such that the combination, 
$Y_{\nu}^{\sf T}Y_{\nu}$, is diagonal. In the case were, $W=1$,
the mixing matrix $U$ is identified as the 
PMNS matrix\footnote{Such an $SU(5)$ model has previously been used to study
$\mu\to e\gamma$, hadronic EDMs, etc. \cite{Hc}}. 
Such a case may occur naturally, 
for example in models with a $U(1)_X$ flavour symmetry.
For simplicity we assumed $W=1$ and that the above matrices are all real.

Throughout the analysis it was assumed that both the
neutrino Yukawa and Majorana matrices have a hierarchical form with
$Y_{\nu_{3}}^{\sf diag}\sim 1$ and $M_{R_{3}}\sim 10^{14}$ GeV, 
as shown in eq.~(\ref{Ynuhier}) and (\ref{MRhier}). We choose such a
hierarchy in the neutrino Yukawa sector with one eye on $b-t-\tau$
Yukawa unification and $SO(10)$ with $\tan \beta \sim 50$.
\bea
&Y_{\nu_{1}}^{\sf diag}<Y_{\nu_{2}}^{\sf diag}<Y_{\nu_{3}}^{\sf diag}
\sim 1&\label{Ynuhier}\\
&M_{R_{1}}^{\sf diag}<M_{R_{2}}^{\sf diag}<M_{R_{3}}^{\sf diag}
\sim 10^{14}\,\,{\rm GeV}&\label{MRhier}.
\eea
This hierarchy results in a neutrino mass ordering which is of the 
Normal Hierarchical type with, $m_{\nu_1}\approx 0.55\times 10^{-6}$ eV,
$m_{\nu_2}\approx 0.009$ eV, $m_{\nu_3}\approx 0.05$ eV. 

The consequence of our choice of $\lambda_{\nu_3}\sim 1$ is that
it shall induce, via RG evolution, large flavour mixing in the 
slepton mass squared matrix. This slepton flavour mixing leads to
a lepton-slepton mis-alignment and produces lepton flavour violation
in the charged lepton sector. The choice of a large neutrino yukawa
coupling enhances this effect and will result in a large rate for 
$\tau \to \mu \gamma$. It should be noted that such a large neutrino
yukawa coupling is inevitable in models based on $SO(10)$.

In the soft SUSY breaking section of the 
Lagrangian flavour blind boundary conditions
are assumed at the GUT scale. 
In this work we are interested in the renormalisation group
effects of off-diagonal elements of the Yukawa couplings as discussed
above. Due to the 
large mixings in the PMNS matrix, these off-diagonal elements can be large
even in the case of GUT scale flavour blindness.
Throughout our analysis 
we have also assumed that
the trilinear coupling, $A_{0}=0$, at the GUT scale. 
Under these assumptions  
the soft Lagrangian becomes flavour blind and hence contains
no new lepton flavour violating sources.

\section{LFV and FCNC Phenomenology}\label{Phenom}

Observations of neutrino oscillations 
imply the existence of massive neutrinos with 
large solar and atmospheric mixing angles. 
The small neutrino masses are most naturally explained via 
the seesaw mechanism with heavy singlet neutrinos.
Even in a basis where both $Y_{e}$ and $M_{R}$ are diagonal
in flavour space $Y_{\nu}$ 
is always left as a possible source of 
flavour violation. 
In SUSY models this flavour violation can be 
communicated to the slepton sector through renormalisation group running. 
The initial communication is from running between the GUT scale and the 
scale of $M_{R}$. Although the scale $M_{R}$ is far above the 
electro-weak scale 
its effects leave a lasting impression on the mass squared matrices of the 
sleptons. Subsequently flavour violation can enter into the charged 
lepton sector through loop diagrams involving the sleptons and 
indeed such effects have been used to predict large branching ratios for 
$\tau \to \mu \gamma$ and $\mu \to e \gamma$ within the MSSM
\cite{KiOl_LFV,BlKi_1,Blazek:2002wq,lfv}. As such, the experimental
bound on charged lepton flavour violation provide important constraints
on the possible flavour mixing of the MSSM and other extensions of the 
SM.

In the Standard Model FCNCs are 
absent at tree-level and only enter at 1-loop order.
In extensions of 
the SM, the MSSM for example, there also exist additional 
sources of FCNC. A clear example comes from the mixings present 
in the squark sector of the MSSM. These mixings will also contribute
to FCNCs at the 1-loop level and could even be larger than their 
SM counterparts. An example that we shall study in this work
is the flavour changing couplings of neutral Higgs bosons 
and the neutral Higgs penguin contribution to such decays
as $B_s^0 \to \mu^+\mu^-$ and $B_s^0\!-\!\bar{B}_s^0$ mixing.
When examining such flavour changing 
in the quark sector it is important to consider the 
constraint imposed by the decay $b\to s\gamma$. 

\subsection{Higgs mediated FCNC}


It has been pointed out that Higgs mediated FCNC processes
could be among the first signals of supersymmetry\cite{Bsmm_papers}. 
In the MSSM radiatively induced couplings between the up Higgs, 
$H_{u}$, and down-type quarks may result in flavour 
changing Higgs couplings. In turn this will lead to large FCNC
decay rates for such process as
$B_s\to\mu^{+}\mu^{-}$ and $B_s^0\!-\!\bar{B}_s^0$ mixing. 
In the Standard Model the 
predicted branching ratio for $B_s\to\mu^{+}\mu^{-}$
is of the order of $10^{-9}$, but in the MSSM such a decay is enhanced
by large $\tan\beta$ and may reach far greater rates.
It is also possible to extend this picture to the charged
lepton sector where similar lepton flavour violating
higgs couplings are radiatively induced. These LFV couplings can
then result in such Higgs mediated decays as $\tau\to 3\mu$, 
the neutral Higgs decays 
$h^{0},\,H^{0},\,A^{0}\to\tau\mu\,$ and even combined with the afore
mentioned FCNC couplings in the decay $B_{s}\to\tau\mu$.

Loop diagrams induce flavour changing couplings of the 
kind, $b^{c}sH_{u}^{0*}$. 
Similar diagrams with Higgs fields replaced by their VEVs 
will also provide down quark mass corrections
and will lead to sizeable 
corrections to the mass eigenvalues \cite{large.dmb} and mixing 
matrices \cite{Blazek:1995nv}. 
Furthermore the 3-point 
function and mass matrix 
can no-longer be simultaneously diagonalised \cite{Hamzaoui:1998nu}
and hence beyond tree-level we shall have flavour changing Higgs
couplings in the mass eigenstate basis. Such flavour changing Higgs
couplings can be summarized as,
\be
    {\mathcal L}_{FCNC} \;=\; 
                            -\ol{d}_{R\,i}
                             \left[X^{S^0}_{RL}\right]_{ij}
                             d_{L\,j} \, S^0
                            -\ol{d}_{L\,i} 
                             \left[X^{S^0}_{LR}\right]_{ij}
                             d_{R\,j} \, S^0\,  .
\label{L_FCNC}
\ee
These flavour changing couplings can 
in fact be related in a simple way to the
finite non-logarithmic mass matrix corrections \cite{Blazek:2003hv},
\be
\left[X^{S^0}_{RL}\right]_{ij} = 
\frac{1}{\sqrt{2}}\frac{1}{c_\beta}
\left(\frac{\delta m_d^{finite}}{v_u}\right)_{ij}\, A_{S^0}
\ee
where, 
$A_{S^0}=\left( s_{\alpha -\beta},\, c_{\alpha - \beta},\,-i   \right)$,
for $S^0= \left( H^0,\,h^0,\,A^0 \right)$.
It is clear that the FCNC couplings are related as, 
$\left[X_{RL}\right]=\left[X_{LR}\right]^{\dagger}$.
In general we should also notice that, 
$\left[X_{RL}\right]_{ij}\approx \frac{m_j}{m_i}\left[X_{RL}\right]_{ji}$.
Hence, in the case of, $(i,\,j)=(b,\,s)$, we have 
$\left[X_{RL}\right]_{bs}\approx \frac{m_s}{m_b}\left[X_{LR}\right]_{bs}$.

In the MSSM with large $\tan\beta$ the dominant contribution 
to $B_s\to\ell^+\ell^-$ comes from the penguin diagram where 
the dilepton pair is produced from a virtual Higgs state
\cite{Bsmm_papers}. 
Thus in combination with the standard tree-level term 
${\mathcal L}_{\ell\ell H} \:=\: -y_\ell \ol{\ell}_R\ell_LH_d^0 
                                 + h.c.$
the dominant $\tan\beta$ enhanced contribution to the 
branching ratio turns out to be,
\bea
  Br(B^0_s\to\mu^+\mu^-) 
                     &=& 
                       1.75\times 10^{-3}\; 
                       \left| 
                             \frac{(\delta m_{d})_{32}^\dagger}
                                  {m_bV_{ts}}
                       \right|^2
                       \;
                       \left[ \frac{V_{ts}}{0.04} \right]^2
                       \left[ \frac{y_{\mu}}{0.0311} \right]^2
                       \left[ \frac{M_{170}}{v_u} \right]^2
                       \left[ \frac{\tan\beta}{50} \right]^2
\nonumber\\ 
              &\times &         
                       \left[
                          \left(
                              \frac{c_\alpha s_{\alpha-\beta}}
                                   {\left( \frac{M_{H^0}}{M_{100}}\right)^2}
                              -
                              \frac{s_\alpha c_{\alpha-\beta}}
                                   {\left( \frac{M_{h^0}}{M_{100}}\right)^2}
                          \right)^2
                          +
                              \frac{s_\beta^2}
                                   {\left( \frac{M_{A^0}}{M_{100}}\right)^4}
                       \right], 
\label{br}
\eea
where the matrix $\delta m_d^\dagger$ is in the 
1-loop mass eigenstate basis.

In the large $\tan\beta$ limit the 
Higgs Double Penguin(DP) contribution
to $B_s^0\!-\! \bar{B}_s^0$ mixing is dominant
\cite{Buras:2002vd}. 
Following the notation of eq.~(\ref{L_FCNC}),
we can write the neutral Higgs contribution to the 
$\Delta B=2$ effective Hamiltonian as,
\bea
{\mathcal H}_{\sf eff}^{\Delta B=2}
&=& 
\frac{1}{2}\sum_{S}\,\frac{[X^S_{RL}]_{bs}[X^S_{RL}]_{bs}}{-M_S^2}\,\,Q_1^{SLL}
+
\frac{1}{2}\sum_{S}\,\frac{[X^S_{LR}]_{bs}[X^S_{LR}]_{bs}}{-M_S^2}\,\,Q_1^{SRR}
\nonumber\\
&&\,\,\,\,
+\sum_{S}\, \frac{[X^S_{RL}]_{bs}[X^S_{LR}]_{bs}}{-M_S^2}\,\,Q_2^{LR}
\label{Heff}
\eea
where we have defined the operators,
\bea
Q_1^{SLL} &=& (\ol{b}P_L s)\,(\ol{b}P_L s)\\
Q_1^{SRR} &=& (\ol{b}P_R s)\,(\ol{b}P_R s)\\
Q_2^{LR}  &=& (\ol{b}P_L s)\,(\ol{b}P_R s)
\label{ops}
\eea
It is common at this point to notice that the contribution to
$Q_2^{LR}$ is dominant over $Q_1^{SLL,SRR}$ due to a suppression
from the sum over Higgs fields, 
${\mathcal F}^-=\left( \frac{s_{\alpha-\beta}^2}{M_H^2}+
\frac{c_{\alpha-\beta}^2}{M_h^2}-
\frac{1}{M_A^2}  \right)$. The contribution to $Q_2^{LR}$ receives
a factor ${\mathcal F}^+=\left( \frac{s_{\alpha-\beta}^2}{M_H^2}+
\frac{c_{\alpha-\beta}^2}{M_h^2}+
\frac{1}{M_A^2}  \right)$ .
It turns out that this typically results in a suppression factor, 
$\frac{-1}{10}\lesssim{\mathcal F}^-/{\mathcal F}^+\lesssim\frac{-1}{25}$. 
Recalling that
$[X_{LR}]_{bs}\sim \frac{1}{40}[X_{RL}]_{bs}$, it may be possible
for the $Q_1^{SLL}$ contribution to give a significant effect. 
On the other hand, the contribution to $Q_1^{SRR}$ is highly
suppressed.

Following the above conventions 
we can write the double penguin contribution as,
\bea
\Delta M_s^{DP}&\equiv& \Delta M_s^{LL}+\Delta M_s^{LR}\nonumber\\
&=&
-\frac{1}{3}M_{B_s}f_{B_s}^2 P_1^{SLL}\, 
\sum_{S}\,
\frac{[X^S_{RL}]_{bs}[X^S_{RL}]_{bs}+[X^S_{LR}]_{bs}[X^S_{LR}]_{bs}}{M_S^2}
\label{dmsDP}\\
&&-\frac{2}{3}M_{B_s}f_{B_s}^2 P_2^{LR}\, 
\sum_{S}\,
\frac{[X^S_{RL}]_{bs}[X^S_{LR}]_{bs}}{M_S^2}
\nonumber
\eea
Here $P_1^{SLL}=-1.06$ and $P_2^{LR}=2.56$, 
include NLO QCD renormalisation group factors \cite{Buras:2002vd}.
After taking into account the relative values of ${\mathcal F^{\pm}}$, 
the two $P$'s and the factor
of 2 in eq.~(\ref{dmsDP}), we can see that there is a relative suppression, 
$\frac{1}{3}\lesssim\Delta M_s^{LL}/\Delta M_s^{LR}\lesssim \frac{4}{5}$.
This relative suppression shall be discussed further in the following
section.

There is a large non-perturbative uncertainty in the 
determination of $f_{B_s}$. Two recent lattice determinations provide
\cite{Hashimoto:2004hn,Gray:2005ad},
\bea
f_{B_s}^{'04}&=&230\pm 30 \,{\rm MeV}\label{fBs1}\\
f_{B_s}^{'05}&=&259\pm 32 \,{\rm MeV},\label{fBs2}
\eea
which in turn give different direct Standard Model predictions for 
$\Delta {M_s^{\rm SM}}$,
\bea
\Delta M_s^{\rm SM'04}&=&17.8\pm 8 {\rm ps}^{-1}\\
\Delta M_s^{\rm SM'05}&=&19.8\pm 5.5 {\rm ps}^{-1}
\label{dmsSM}
\eea
The recent precise Tevatron measurement of $\Delta M_s$ 
is consistent with
these direct SM prediction but with a lower central value 
\cite{cdf:DMs,Abazov:2006dm} ,
\be
\Delta M_s^{\rm CDF}=17.31^{+0.33}_{-0.18}\pm 0.07 {\rm ps}^{-1}
\label{dmsCDF}
\ee


\subsection{Higgs mediated lepton flavour violation}

We have seen that flavour changing 
can appear in the couplings of the neutral Higgs bosons and is enhanced by 
large $\tan \beta$. In the quark sector, interactions of the form, 
$\bar{d}_{R}d_{L}H_{u}^{0*}$, are generated at one-loop and 
at large $\tan \beta$ can become comparable to the tree-level interaction, 
$\bar{d}_{R}d_{L}H_{d}^{0}$. 
Similar Higgs-mediated flavour violation can also occur in 
the lepton sector of SUSY seesaw 
models through interactions of the form $\bar{e}_{R}e_{L}H_{u}^{0*}$.
This leads to the possibility of large branching ratios for 
Higgs-mediated LFV processes such as 
$B_{s}\to \tau \mu$, $\tau\to 3 \mu$ and lepton flavour violating 
Higgs decays \cite{Dedes:2002rh,Babu:2002et,LFVhiggs}.

As we did for the down-quarks, 
we can find effective flavour changing Higgs 
vertices induced after heavy sparticles are integrated out of the 
lagrangian. In the 1-loop corrected eigenstate basis these couplings
are again related to the charged lepton finite mass corrections, 
$(\delta m_e)^{\sf finite}_{ij}/v_u$.


Possibly the most interesting application of such lepton 
flavour violating couplings is in the decays of MSSM Higgs
bosons. 
These lepton flavour violating Higgs couplings shall
facilitate flavour violating Higgs decays. We can write the 
partial widths of the lepton flavour violating 
Higgs boson decays within the MSSM as follows,
\bea
\Gamma_{S^{0}\to \ell_i \ell_j}\!\!\!&=&\!\!\!
\frac{1}{16\pi}\frac{(\delta m_e)_{ij}^2}{v_u^2} 
\;\left|a^{S}\right|^2 M_{S}
 \left(1-x_i-x_j \right)
\sqrt{(1-(x_i+x_j)^2)(1-(x_i-x_j)^2)}
\label{width_Htm}\\
&&a^{S}=\left[\;\sin(\alpha - \beta),\;\cos(\alpha - \beta),\;  -i\;\right]
\hspace{4mm}{\rm for}\hspace{4mm}S=\left[\;H^{0},\;h^{0},\;A^{0}\;\right].
\nonumber
\eea
Here, $S^{0}=H^{0},h^{0},A^{0}$ represent the three physical Higgs states,
$M_{S}=M_{H},M_{h},M_{A}$ are their masses, 
$\ell_i=\tau,\mu, e$ are the three charged 
lepton states and $x_i=(m_{\ell_i}/M_{S})^2$.


We can also make use of the LFV Higgs couplings to study the 
Higgs mediated contributions to the process $\tau \to 3\mu$ for example. 
The dominant Higgs contributions will come from the 
Higgs penguin contribution.
In the MSSM with large $\tan\beta$ the dominant contribution to 
the branching ratio of $\tau \to 3\mu$ turns out to be,
\bea
\hspace{-0.5cm}{\rm Br}(\tau \to 3 \mu)
= \frac{\tau_{\tau}}{4096\pi^{3}}m^{5}_{\tau}
\left(
\frac{(\delta m_{e})_{23}}{v_{u}}\frac{\lambda_{\mu}}{2c_{\beta}}
\right)^{2}
\left[
\left(
\frac{c_{\alpha}s_{\alpha-\beta}}{M_{H^{0}}^{2}}
-
\frac{s_{\alpha}c_{\alpha-\beta}}{M_{h^{0}}^{2}}
\right)^{2}
+
\left(
\frac{s_{\beta}}{M_{A^{0}}^{2}}
\right)^{2}
\right]
.\label{LFV:eq11}
\eea
Here $\tau_{\tau}$ is the lifetime of the tau lepton 
and $\lambda_{\mu}$ is the 
Yukawa coupling of the muon.
The present experimental bound for this decay is as follows
\cite{Aubert:2003pc},
\bea
{\rm BR}(\tau\to 3\mu)&<&1.9 \times 10^{-7}\,\,\, {\rm at\,\,90\%\,\,C.L.}
\label{t3m}
\eea

Going one step further, 
the lepton flavour violating Higgs couplings
can also be combined 
with the quark flavour changing coupling studied earlier.
In this way we can also study the LFV and FCNC process $B_{s}\to \tau\mu$.
In the MSSM with large $\tan\beta$ 
the dominant Higgs contribution will again come from the penguin diagram 
mediated by Higgs bosons. The branching
ratio for this decay may be written as,
\bea
{\rm Br}(B_{s} \to \tau^{+}\mu^{-})
&\hspace{-0.2cm}=&\hspace{-0.2cm} 
\frac{\tau_{B_{s}}}{256\pi}
\frac{(\delta m_{b})_{23}^{2}}{v_{u}^{2}}
\frac{(\delta m_{e})_{23}^{2}}{v_{u}^{2}}
\frac{1}{c_{\beta}^{4}}f_{B_{s}}^{2}\frac{M_{B}^{5}}{m_{b}^{2}}
\nonumber\\
&&\hspace{-3.5cm}\times
\left[
\frac{s^{2}_{\alpha-\beta}}{M_{H^{0}}^{2}}
+
\frac{c^{2}_{\alpha-\beta}}{M_{h^{0}}^{2}}
+
\frac{1}{M_{A^{0}}^{2}}
\right]^{2}
(1-x_{\mu}-x_{\tau})\sqrt{1-2(x_{\tau}+x_{\mu})+(x_{\tau}-x_{\mu})^{2}}
,\label{LFV:eq12}
\eea
here $x_{i}=(m_{\ell_i}/M_{B_{s}})^{2}$.

For the analysis two-loop RGEs for the dimensionless couplings and 
one-loop RGEs for the dimensionful couplings were used to 
run all couplings down to the scale $M_{3R}$ where the heaviest
right-handed neutrino is decoupled from the RGEs. Similar steps
were taken for the lighter $M_{2R}$ and $M_{1R}$ scales, 
and finally with all three right-handed neutrinos decoupled
the solutions for the MSSM couplings and spectra were computed 
at the $Z$ scale. This includes full one loop SUSY threshold corrections 
to the fermion mass matrices and all Higgs masses
while the sparticle masses are obtained at tree level.
A detailed description of the numerical procedure can also be found
in reference \cite{Parry:2005fp}.
A $\chi^2$ function $\;\sum_i\;(X_i^{th}-X_i^{exp})^2/\sigma_i^2\;$
is evaluated based on the agreement between the theoretical predictions
and 24 experimental observables.

The analysis uses approximately 100 evenly spaced points in the SUSY 
parameter space each with values of $m_{0}$ and $M_{1/2}$ in the 
plane, $m_{0}=300 - 1400$ GeV and $M_{1/2}=300-1000$ GeV. For each of the 
100 points we also held fixed $\mu=+120$ GeV, $\tan\beta=50$
and $A_{0}=0$. The remaining
input parameters are allowed to vary
in order to find the minimum of our $\chi^2$ function.

\section{Discussion}\label{results}

In this section we shall present the results of our analysis.
We shall discuss the LFV and FCNC phenomenology is turn.

\begin{figure}[ht]
\begin{center}
\rotatebox{-90}{\scalebox{0.27}{\includegraphics*{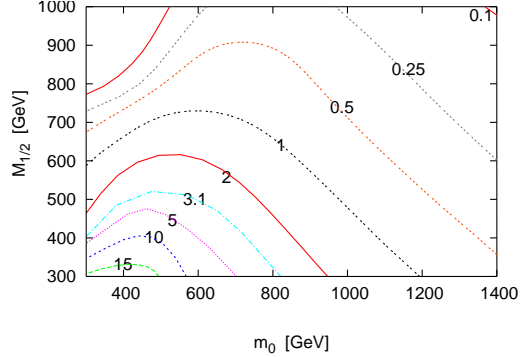}}}
\begin{minipage}[t]{15cm}
\caption{\small{ A contour plot of Br$(\tau\to\mu\gamma)\times 10^{7}$
in the $m_{0}-M_{1/2}$ plane with $\mu=120$ GeV and $A_{0}=0$. 
The present experimental bound is marked by 
the contour labelled "3.1".}\label{plot_tmg}}
\end{minipage}
\end{center}
\end{figure}
Let us firstly look at the important lepton 
flavour violating decay $\ell_i\to\ell_j\gamma$.
A contour plot showing the variation of the branching
ratio for $\tau\to\mu\gamma$ in the $m_{0}-M_{1/2}$ plane 
is given in fig.~{\ref{plot_tmg}}. This plot shows that the 
rate for this process is very high over most of the plane
and even in the low $m_{0}-M_{1/2}$ region the rate can exceed
the present experimental upper bound. 
These large rates are a
direct consequence of our choice of $Y_{\nu_3}^{\sf diag}\sim 1$. 

\begin{figure}[ht]
\begin{center}
\rotatebox{0}{\scalebox{0.33}{\includegraphics*{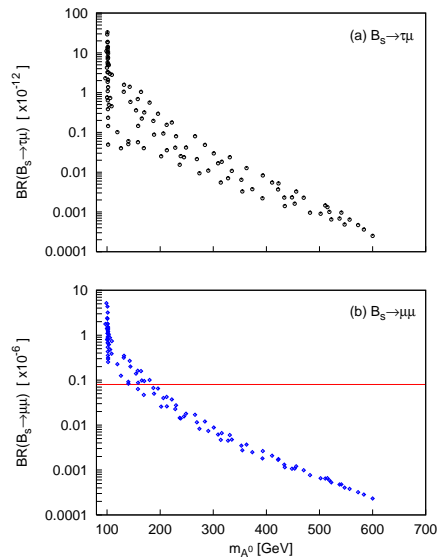}}}
\vskip-15mm
\begin{minipage}[t]{15cm}
\caption{\small{ The Higgs mediated contribution to the 
branching ratio for $B_{s}\to \mu\mu(\tau\mu)$ 
is plotted in the lower(upper) panel against the Pseudoscalar 
Higgs mass, $m_{A^{0}}$. The horizontal line at $0.8\times 10^{-7}$ shows 
the present experimental upper bound. 
}}\label{plot_Bs}
\end{minipage}
\end{center}
\end{figure}
The plot for the Higgs mediated contribution to the flavour
changing neutral current decay, $B_s\to\mu^+\mu^-$, is presented in
the lower panel of fig.~\ref{plot_Bs}.
As previously reported, the branching ratio for 
$B_s\to\mu^+\mu^-$ is particularly interesting
with rates ranging from $10^{-10}$ for heavy $m_{A^0}$ up to
almost $10^{-5}$ for a particularly light $m_{A^0}$. The present 
$90\%$ C.L. experimental upper bound of 
$0.8\times 10^{-7}$\cite{cdf:Bsmm}
excludes the highest predicted rates and hence is beginning to 
probe the Higgs sector into the region $m_{A^0}\sim 150$ GeV. This 
rare decay is certainly of particular interest and future studies 
will continue to probe the Higgs sector to an even greater extent.
Within the standard model the expectation is that 
Br$(B_s\to\mu^+\mu^-)_{SM}\sim 10^{-9}$. Therefore we can see 
from fig.~\ref{plot_Bs} that the Higgs mediated contribution
to this process would dominate as long as $m_{A^{0}}< 500$ GeV
and this decay has the possibility of being the first 
indirect signal of supersymmetry.

\begin{figure}[ht]
\begin{center}
\rotatebox{-90}{\scalebox{0.225}{\includegraphics*{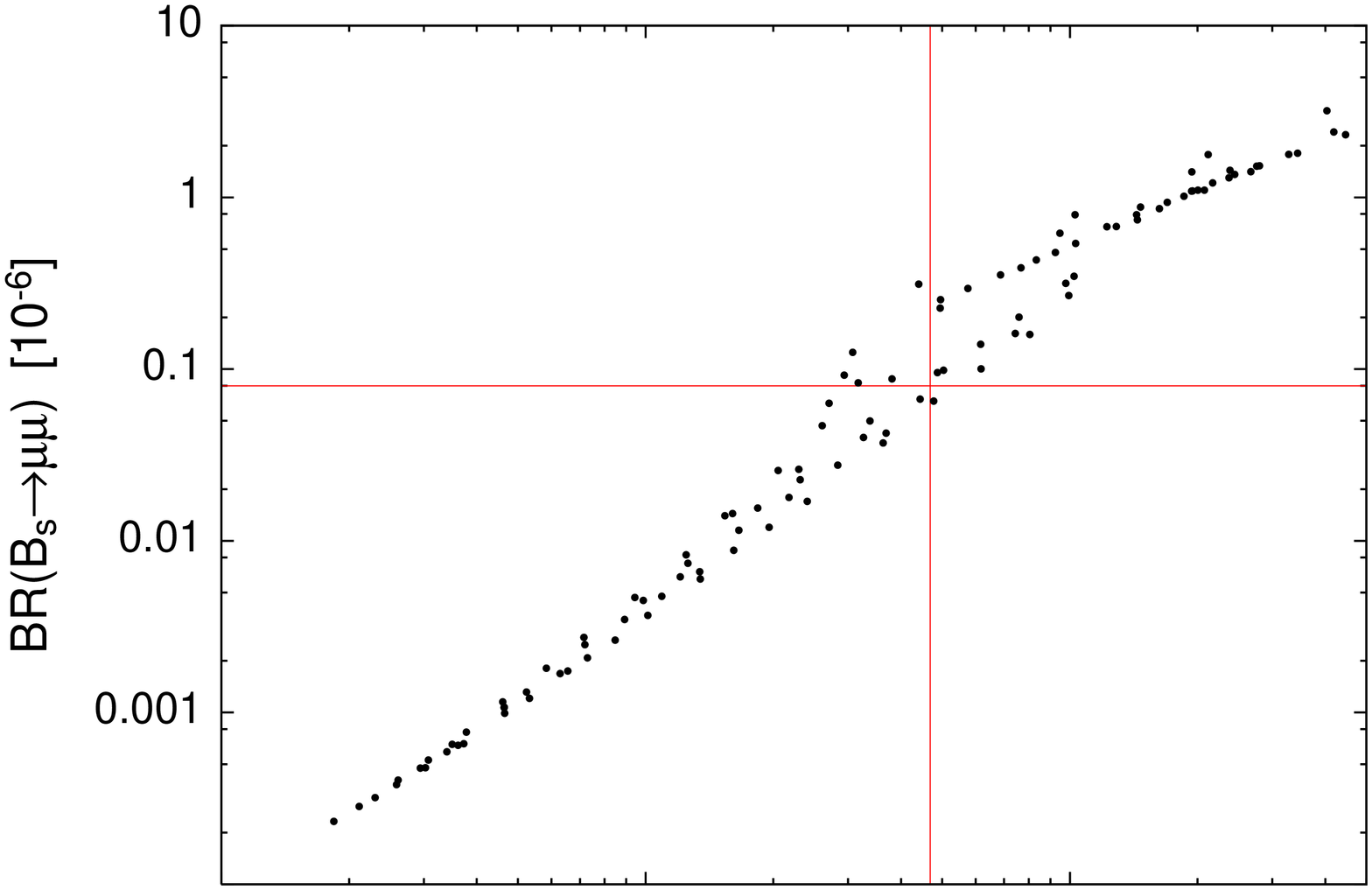}}}\\
\vskip-5.7mm
\rotatebox{-90}{\scalebox{0.225}{\includegraphics*{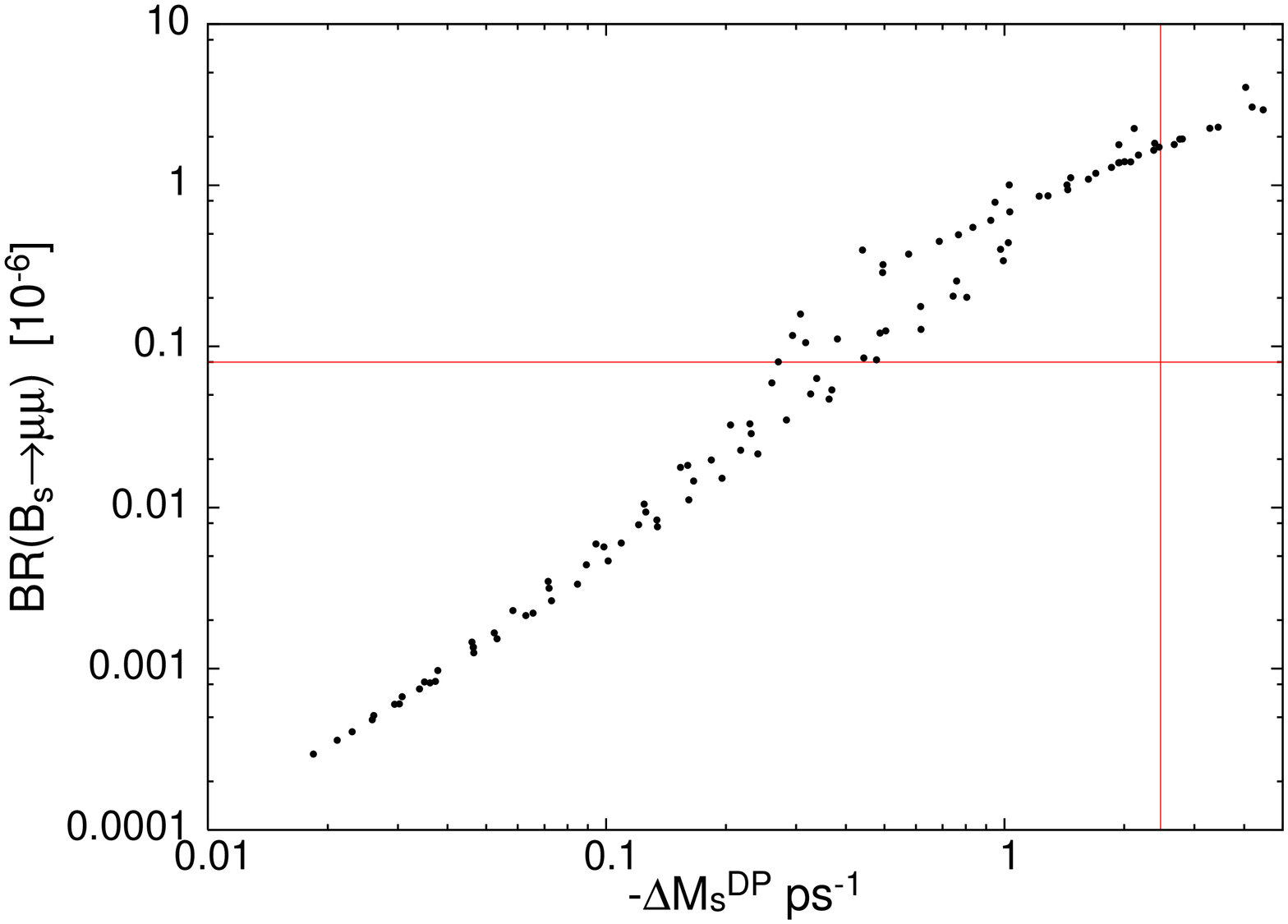}}}\\
\begin{minipage}[t]{15cm}
\caption{\small{ 
The correlation of Br$(B_s\to\mu^+\mu^-)$ 
and $\Delta M_s^{\rm DP}$ are given for $f_{B_s}=230$ MeV
(upper panel) and $f_{B_s}=259$ MeV(lower panel). 
The horizontal line in both panels shows the present experimental
bound on Br$(B_s\to\mu^+\mu^-)$. The vertical lines mark the central
value of $(\Delta M_s^{\rm CDF}-\Delta M_s^{SM})$.
}}\label{plot_DMs}
\end{minipage}
\end{center}
\end{figure}
The lepton flavour violating decay $B_s\to\tau\mu$, plotted 
against the Pseudoscalar Higgs mass, is shown in 
the upper panel of fig.~\ref{plot_Bs}.
This plot shows that the branching ratio for this decay 
could be as large as $\sim 10^{-10}$, but are certainly
not as interesting as the flavour conserving decay just discussed. 
The branching ratio
for the decay $\tau\to 3\mu$ are of a similar order of magnitude 
to those in the upper panel of fig.~\ref{plot_Bs}. 

In the limit of large $\tan\beta$,
$B_s\to\mu^+\mu^-$ and $\Delta M_s$ are
correlated. This correlation is shown in the two panels
of fig.~\ref{plot_DMs}. For these two panels the two different
values of $f_{B_s}$ listed in eq.~(\ref{fBs1},\ref{fBs2}) 
are used. The upper panel
($f_{B_s}=230$ MeV) shows that the central value of 
$(\Delta M_s^{\rm CDF}-\Delta M_s^{\rm SM})$ coincides
with the bound from Br$(B_s\to\mu^+\mu^-)$.
The lower panel ($f_{B_s}=259$ MeV) shows that the data points
with $\Delta M_s^{\rm DP}$ at the central value, are in fact
ruled out by the bound on Br$(B_s\to\mu^+\mu^-)$. The uncertainty 
in the SM prediction
for $\Delta M_s$ is rather large and in fact all of the data points
of fig.~\ref{plot_DMs} are allowed by the recent tevatron measurement.
These two panels clearly show that the interpretation of the 
recent measurement depends crucially on the uncertainty in the 
determination of $f_{B_s}$. 

\begin{figure}[ht]
\begin{center}
\rotatebox{-90}{\scalebox{0.25}{\includegraphics*{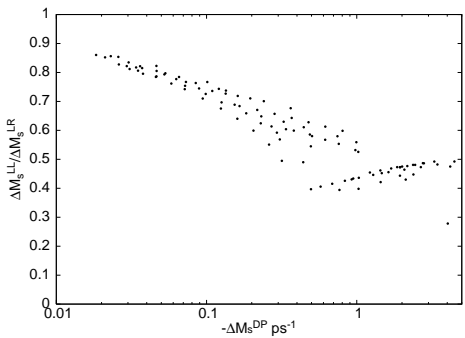}}}
\begin{minipage}[t]{15cm}
\caption{\small{Plot for the ratio, $\Delta M_s^{LL}/\Delta M_s^{LR}$,
of the Higgs contribution to the 
operators $Q_1^{SLL}$ and $Q_2^{LR}$.
}}\label{plot_DMsLLRL}
\end{minipage}
\end{center}
\end{figure}

The plot in  fig.~\ref{plot_DMsLLRL} shows the ratio, 
$\Delta M_s^{LL}/\Delta M_s^{LR}$, 
of the contributions to 
the operators $Q_1^{\rm SLL}$ and $Q_2^{\rm LR}$
as defined in eq.~(\ref{dmsDP}). 
It is commonly assumed that the contribution to the $Q_2^{\rm LR}$
operator, $\Delta M_s^{LR}$, is dominant. From fig.~\ref{plot_DMsLLRL}
we can see that the contribution of the $Q_1^{\rm SLL}$ operator, 
$\Delta M_s^{LL}$, is 
between 40\% and 90\% of $\Delta M_s^{LR}$ and 
hence is significant.

\begin{figure}[ht]
\begin{center}
\rotatebox{0}{\scalebox{0.33}{\includegraphics*{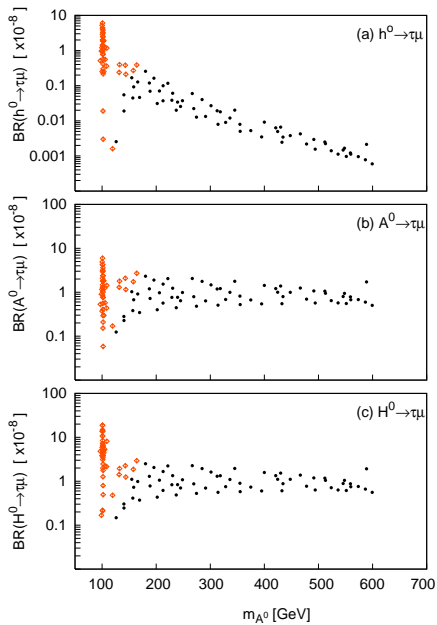}}}
\vskip-5mm
\begin{minipage}[t]{15cm}
\caption{\small{ BR($H^{0}\to \tau\mu$) plotted against the Pseudoscalar
Higgs mass, $m_{A^{0}}$. 
For each of these fitted points we have assumed the following
values, $\mu=120$ GeV and $A_{0}=0$. Here the dots(circles) 
mark points for which $Br(B_s\to\mu^+\mu^-)$ is below(above) the present
experimental limit.
}}\label{plot_Htm}
\end{minipage}
\end{center}
\end{figure}

Our numerical results for the branching ratio of $H^{0}\to\tau\mu$
are presented in fig.~\ref{plot_Htm}c. The rates for this process
are plotted against the Pseudoscalar higgs mass, $m_{A^{0}}$. 
The predicted rates range from $10^{-9}$ up to a few times $10^{-7}$.
In fig.~\ref{plot_Htm}c we can see a broad band of points stretching
from $m_{A^{0}}=100$ GeV up to $m_{A^{0}}=600$ GeV showing that most 
points are predicting a rate of around $10^{-8}$. 
The decay rate appears to be 
almost independent of the Pseudoscalar higgs mass, although
there is a slight peak at the lower range of the higgs mass, which is 
where the highest rates are achieved.
Notice that the data points shown in fig.~\ref{plot_Htm} 
are divided into two groups. 
The grouping is determined by whether
the $B_s\to\mu^+\mu^-$ bound is in excess or not, as indicated in the 
figure caption. From this grouping we can see that the points with the
highest predicted $H^0\to\tau\mu$ rates appear to be excluded by the 
$B_s\to\mu^+\mu^-$ bound. In this way the 
Higgs mediated contribution to the $B_s$ decay can provide 
additional information on the allowed Higgs decay rate.

Fig.~\ref{plot_Htm}b shows our predictions 
for the lepton flavour violating Pseudoscalar
higgs decay, $A^{0}\to\tau\mu$. The rates 
for the decay of the Pseudoscalar are 
almost identical to those of the heavy
CP-even higgs shown in fig.~\ref{plot_Htm}c.

The rate for the decay of the lightest 
higgs boson is shown in fig.~\ref{plot_Htm}a where
it is again plotted against the Pseudoscalar higgs
mass. The predicted rates for this decay 
show a very different dependence 
upon $m_{A^{0}}$, as they are spread over a large 
range from $10^{-11}$ to $10^{-7}$. 
The branching ratio for the 
lightest higgs appears to be inversely proportional to the 
Pseudoscalar higgs mass. 
Hence this LFV decay will only be interesting
if $m_{A^{0}}< 300$ GeV where its rate can be comparable to those
for the other neutral higgs states. The data points for the lightest
Higgs boson decay of fig.~\ref{plot_Htm}a have again been divided
into two groups. As before, the two groupings 
depend upon whether the $B_s\to\mu^+\mu^-$ bound is being exceeded or not.
The plot shows that the data points for particularly light Pseudoscalar
Higgs mass are excluded by the $B_s\to\mu^+\mu^-$ bound. These excluded
points correspond to the largest predictions for the decay $h^0\to\tau\mu$
and leaves, $3\times 10^{-9}$, the highest allow decay rate.

\section{Conclusions}\label{conc}

We have analysed a supersymmetric-seesaw model constrained by
$SU(5)$ unification at the GUT scale.
We have been concerned with making predictions for lepton flavour
violating decay processes such as $\tau\to\mu\gamma$, $\mu\to e\gamma$
and $h^{0},\,H^{0},\,A^{0}\to\tau\mu$. With lepton flavour violation in mind, 
we chose to study the large $\tan\beta$ 
region of parameter space where such effects are enhanced.
In addition we chose $Y_{\nu_3}^{\sf diag}\sim 1$ under the influence of 
$SO(10)$ unification in which $\tan\beta$ is naturally large.

Our numerical procedure utilises a complete top-down global 
$\chi^2$ fit to 24 low-energy observables. Through this electroweak fit
we were able to analyse lepton flavour violating 
rates for rare charged lepton decays. The choice of a large third family 
neutrino Yukawa coupling naturally leads to large 23 lepton flavour
violation. Within this scenario we have also 
used these numerical fits to analyse the
FCNC processes $B_s\to\mu\mu(\tau\mu)$,
$B_s^0\!-\!\bar{B}_s^0$ mixing  and the 
lepton flavour violating decays of MSSM Higgs bosons. As a result,
we have been about to make realistic predictions for this phenomenology
while ensuring $\ell_i\to\ell_j\gamma$ and $b\to s \gamma$ constraints
are satisfied.

Our model predicts a $\tau\to\mu\gamma$ 
rate near the present experimental limit.
We have seen that 
the branching ratio for $H^{0}\to\tau\mu$ could be
interesting with a rate as high as few $10^{-7}$. 
The Higgs mediated contribution to the 
branching ratio for $B_s\to\mu^+\mu^-$ was found to be particularly
large and could even exceed the current experimental bound.
In addition to the constraint of $\tau\to\mu\gamma$, we also 
saw that the $B_s\to\mu^+\mu^-$
bound may also act as a stringent restriction on 
the allowed rates for $\phi^0\to\tau\mu$. 
We found the rates for the LFV decays $B_s\to\tau\mu$
and $\tau\to 3\mu$ are rather small $< 10^{-10}$. 

The correlation of $B_s\to\mu^+\mu^-$ and $\Delta M_s$ in the
large $\tan\beta$ limit was also studied. The constraint from the 
recent tevatron measurement is highly dependant upon the 
determination of $f_{B_s}$. It was also found that the Higgs 
contribution to the operator $Q_1^{\rm SLL}$ can compete with 
that of the $Q_2^{\rm LR}$ operator.

\section*{Acknowledgements}
The author would like to thank C.H. Chen for interesting discussions
and for hospitality during the miniworkshop.



\end{spacing}

\end{document}